\documentclass[aps, twocolumn,superscriptaddress]{revtex4-2} \usepackage{natbib}
\usepackage{import}
\usepackage{xifthen}
\usepackage{transparent}
\usepackage[colorlinks=true]{hyperref}
\usepackage{array}
\usepackage{tikz}
\usetikzlibrary{decorations.pathmorphing,arrows.meta,positioning,shapes.multipart,calc}

\usetikzlibrary{arrows.meta,decorations.markings}

\tikzset{
  ln/.style={line width=0.9pt},
  midarrow/.style={postaction={decorate},
    decoration={markings,mark=at position 0.7 with {\arrow{Latex}}}},
  lab/.style={fill=white,inner sep=1pt,font=\footnotesize},
  pillar/.style={rounded corners=2pt,fill=gray!60,draw=none}
}

\usepackage{verbatim, bm}
\usepackage{physics}
\usepackage{textcomp}
\usepackage{nicefrac}
\usepackage{mathrsfs}  
\usepackage[normalem]{ulem}
\usepackage{mathtools}
\usepackage{amsfonts}
\usepackage{amssymb}
\usepackage{color}
\usepackage{amsthm}
\usepackage{slashed}
\usepackage{amsmath}
\usepackage{wrapfig}
\usepackage{dsfont}
\usepackage{esint}

\hypersetup{
    unicode=false,          
    pdftoolbar=true,        
    pdfmenubar=true,        
    pdffitwindow=false,     
    pdfstartview={FitH},    
    pdfsubject={},          
    pdfcreator={},          
    pdfproducer={},         
    pdfkeywords={} {} {},   
    pdfnewwindow=true,      
    colorlinks=true,        
    linkcolor=magenta,      
    citecolor=blue,         
    filecolor=magenta,      
    urlcolor=blue           
} 
\allowdisplaybreaks

\renewcommand{\bar}{\overline}
\renewcommand{\tilde}{\widetilde}

\newcommand{\UU}{\operatorname{U}}

\newcommand{\id}{\mathds{1}}

\begin{document}
\title{Twisted quantum doubles are sign problem-free}
\author{Leyna Shackleton}
\affiliation{Department of Physics, Massachusetts Institute of Technology, Cambridge MA 02139, USA}
\email{lshackle@mit.edu}
\date{\today\vspace{0.4in}}

\begin{abstract}
  The sign problem is one of the central obstacles to efficiently simulating quantum many-body systems. It is commonly believed that some phases of matter, such as the double semion model, have an intrinsic sign problem and can never be realized in a local sign problem-free Hamiltonian due to the non-positivity of the wavefunction. We show that this is not the case. Despite not being stoquastic, the standard criteria for the existence of a sign problem, the double semion model, and all twisted quantum double phases of matter for finite groups $\mathcal{G}$ can be realized in local Hamiltonians that are sign problem-free within a stochastic series expansion. The lack of a sign problem is not fine-tuned and does not require the Hamiltonian to be exactly solvable, with sign problem-free perturbations allowing access to a variety of topological phase transitions.
\end{abstract}

\maketitle
\section{Introduction}
Large-scale simulations of interacting quantum systems are central to understanding the physical world. However, the exponential growth in Hilbert space dimension with system size causes many models to be intractable numerically. Quantum Monte Carlo (QMC) methods~\cite{sandvik1991, becca2017} cure this growth in complexity by stochastic sampling, and have offered crucial insight into complex phenomena such as critical Fermi surfaces~\cite{berg2012}, deconfined criticality~\cite{sandvik2007}, and confinement in gauge theories~\cite{xu2019a}. The \textit{sign problem} reflects an obstacle for QMC methods due to the inability to decompose the quantum partition function as a sum of positive probabilities. The sign problem is present in many interesting models in condensed matter physics, including the electron Hubbard model, frustrated antiferromagnets, and interacting electron gases.

A natural question is whether a phase can possess an \textit{intrinsic} obstacle for realization in a sign problem-free (SPF) system.
Any Hamiltonian is SPF in its energy eigenbasis, so claims of intrinsicness require additional assumptions. \textit{Stoquasticity} and \textit{wavefunction non-positivity} are often used as a proxy for an intrinsic sign problem in bosonic systems~\cite{bravyi2007}. Stoquasticity demands that all off-diagonal matrix elements of a Hamiltonian are non-positive, and is a sufficient condition for a model to be SPF. By the Perron-Frobenius theorem, the ground state wavefunction can be chosen to have all positive entries in the stoquastic basis. Consequently, if a phase necessarily has negative ground-state amplitudes in \textit{every} local basis, no local stoquastic representation exists.

Intrinsic wavefunction non-positivity in local bases has long been used to test for an intrinsic sign problem. Intrinsic wavefunction non-positivity was first proven for the double semion phase~\cite{hastings2016} and was later generalized to broad classes of topological quantum field theories (TQFTs) and chiral matter~\cite{ringel2017,smith2020,golan2020, kim2022, seo2025a}. This criterion is stringent: a recent study~\cite{seo2025a} finds intrinsic wavefunction non-positivity in 398 out of 405 bosonic topological orders up to rank 12.

It is important to critically examine purported sign problems as they may turn out to be easily cured or even non-existent. An example of this is the spinless $t-V$ model on bipartite lattices, which can be simulated using fermion bag methods~\cite{huffman2014}. Despite stoquasticity not being a necessary condition for a SPF Hamiltonian in widely-used QMC schemes such as the stochastic series expansion~\cite{sandvik1991}, no counterexamples to these no-go arguments have appeared. Recent studies~\cite{hen2019,gupta2019, hen2021} show that a weaker condition is sufficient: a Hamiltonian is SPF if it can be made stoquastic by rotating each many-body basis state by a state-dependent phase. This refined definition does not require actually implementing these phase rotations; it is merely a more accurate criteria for detecting when a sign problem will appear in these QMC schemes. This observation questions the relevance of wavefunction non-positivity: since any wavefunction can trivially be made positive by such phase redefinitions, the condition of stoquasticity modulo these transformations is essentially orthogonal to the condition of wavefunction non-positivity. Moreover, it raises an important question: does this refined criteria render previously-inaccessible phases of matter accessible?
In this work, we answer this question in the affirmative. We demonstrate that all twisted quantum doubles of a finite group $\mathcal{G}$, including the double semion model, admit local SPF models despite their non-stoquasticity. Our starting point is an analysis of the sign problem in symmetry-protected topological (SPT) phases~\cite{chen2013a,senthil2015}. We show that all bosonic SPTs classified by the group cohomology of a discrete unitary symmetry $\mathcal{G}$ are related to stoquastic Hamiltonians by many-body phase rotations. Thus, these bosonic SPTs have a local SPF, but generically non-stoquastic, Hamiltonian. Gauging then produces topologically ordered models described by twisted quantum doubles, also known as Dijkgraaf-Witten (DW) theories~\cite{dijkgraaf1990}. After gauging, the new Hamiltonian remains related to a stoquastic untwisted quantum double via a non-local but diagonal unitary transformation, rendering them SPF. This includes the aforementioned double semion model and many non-Abelian TQFTs. Neither the SPT models nor their gauged counterparts are fine-tuned and can be perturbed without spoiling the absence of a sign problem. 

The existence of a non-local unitary transformation between twisted and untwisted quantum double ground states is known and has been discussed in tensor network studies~\cite{huang2016a, xu2018, zhang2020c, haller2023}; however, its connection to QMC has not been appreciated. Moreover, a unitary mapping between ground states does not necessarily extend to the full Hamiltonian.


\section{Review of the sign problem beyond stoquasticity}
\label{sec:sign}We summarize the alternative to stoquasticity developed in~\cite{hen2019,gupta2019,hen2021}, which is applicable to bosonic systems studied using a stochastic series expansion (SSE)~\cite{sandvik1991}.

Let $H = \sum_b H_b$ be a bosonic Hamiltonian defined in a local computational basis $\{\ket{z_j}\}$. Each local term $H_b$ maps a basis state either to zero or to another basis state, i.e., acts as a (weighted) permutation on the basis with possibly zero coefficients.

In an SSE simulation, we expand the partition function:
\begin{equation}
  \begin{aligned}
    Z &= \Tr\left[e^{-\beta H}\right] 
    \\
    &= \sum_{n=0}^\infty \sum_{j} \sum_{\{b_1, b_2 \ldots b_n\}} \frac{(-\beta)^n}{n!} \bra{z_j} H_{b_1}H_{b_2} \ldots H_{b_n}\ket{z_j}\,.
    \label{eq:sse}
  \end{aligned}
\end{equation}
We obtain a probabilistic interpretation if each term is nonnegative, which enables a stochastic sampling of Eq.~\ref{eq:sse}. Term-wise stoquasticity, i.e. $\bra{z_j}H_b\ket{z_k} < 0$, satisfies this condition. Since we can always shift $H$ by a constant, this restriction is only non-trivial for $j \neq k$. However, it was emphasized in~\cite{hen2019, gupta2019, hen2021} that each term may individually possesses a non-trivial phase factor, but that these phase factors always cancel out for any non-zero matrix element $\bra{z_j} H_{b_1}H_{b_2} \ldots H_{b_n}\ket{z_j}$. 
This weaker criterion is harder to verify, but progress can be made by noting that these global phases are invariant under $\ket{z_j} \rightarrow e^{i \theta_j} \ket{z_j}$. It has been shown~\cite{hen2021} that a Hamiltonian is SPF under this method iff it is related to a term-wise stoquastic Hamiltonian via these many-body basis rotations. Crucially, only the \textit{existence} of such a transformation is required to ensure that the sign problem vanishes in the original basis. No knowledge of basis rotations is needed to actually perform the QMC. The goal of the following two sections is to show that such a many-body phase rotation is capable of mapping a twisted quantum double to its untwisted, stoquastic, counterpart. While some aspects of these basis rotations are technical, we emphasize that their complexity is of no relevance to an actual QMC practitioner.
\section{Discrete bosonic SPTs are sign problem-free}
As an initial result, we demonstrate that all bosonic SPT states classified by the group cohomology of any discrete unitary symmetry $\mathcal{G}$ are SPF. Since the ground states can always be deformed to a product state by finite-depth local unitary (FDLU) gates, one may argue that no sign problem exists, as has previously been noted~\cite{ellison2021} - this relation has been used to effectively study the Levin-Gu SPT~\cite{dupont2021, dupont2021a} with QMC. Our statement is stronger: these FDLU gates implement many-body basis rotations that preserve the sign structure of QMC. Hence, no transformation to a trivial phase is necessary - all bosonic SPT phases possess a SPF but possibly non-stoquastic Hamiltonian in their original basis.

To start with an explicit example, we consider the Levin-Gu model~\cite{levin2012}, an SPT in $d=2$ with $\mathbb{Z}_2$ symmetry, defined with qubits on a triangular lattice: \begin{equation}
  \begin{aligned}
  H_{\text{Levin-Gu}} = -\sum_j \sigma^x_j \prod_{\triangle_{j k l}} i^{\frac{1}{2} (-1 - \sigma^z_k \sigma^z_l)} \equiv - \sum_j A_j \,.
  \end{aligned}
\end{equation}
Here $\prod_{\triangle_{jkl}}$ runs over all triangles whose vertices include $j$.
This Hamiltonian is non-stoquastic, yet all phase factors cancel out in an SSE. Acting on a basis state, $A_j$ contributes $-1$ if it changes the parity of domain walls and $+1$ otherwise. Returning to the original state requires an even number of parity changes, so every term appearing in Eq.~\ref{eq:sse} is positive. Alternatively, a many-body basis rotation $U = (-1)^{N_{dw}}$, with $N_{dw}$ the domain-wall count, maps $H_{\text{Levin-Gu}}$ to the Ising paramagnet $H_{\text{trivial}} = - \sum_j \sigma_j^x$~\cite{levin2012}, which guarantees that the sign problem of the former vanishes. 

We now state the more general case, although we stress that the prior example will suffice for demonstrating the sign problem-free nature of the double semion model. The idea is the following. For every discrete SPT, there exists a canonical Hamiltonian that can be mapped to a trivial stoquastic ``paramagnet'' via explicitly-known FDLU gates~\cite{chen2013a}. The fact that has not previously been emphasized is that in the standard computational basis, these FDLU gates are precisely the many-body basis rotations that leave the sign structure of an SSE simulation invariant. Hence, these SPT Hamiltonians do not possess a sign problem in their original basis.

We make this argument precise here. For a symmetry group $\mathcal{G}$ in $d$-dimensions, we take our lattice to be a triangulation of $d$-dimensional space. The Hilbert space on each lattice site is spanned by basis states labeled by group elements $g \in \mathcal{G}$. Different SPT phases are classified by the cohomology group $H^{d+1}( \mathcal{G}, \UU(1))$. Given a homogeneous cocycle $\nu_{d+1} \in H^{d+1}(\mathcal{G}, \UU(1))$ which assigns a phase to every collection of $d+2$ group elements, the SPT wavefunction is given by~\cite{chen2013a}
\begin{equation}
  \begin{aligned}
    \ket{\psi} = \sum_{\{g_i\}} \left[ \prod_{\Delta^d} \nu_{d+1}(0, g_{i_1},g_{i_2},\ldots g_{i_{d+1}})^{\epsilon(\Delta^d)} \right] \ket{\{g_i\}}\,,
  \end{aligned}
\end{equation}
where $\epsilon = \pm 1$ specifies the orientation of the cocycle.
This state generalizes a trivial ``paramagnet', $\sum_{\{ g_i\}} \ket{\{g_i\}}$ by attaching phase factors to each basis state. These phase factors are determined by evaluating the cocyle $\nu_{d+1}(0, g_{i_1},g_{i_2},\ldots g_{i_{d+1}})$ on each local simplex. Different elements of the cohomology group classify topologically distinct ways of assigning these phase factors.
In other words, it is equivalent to a trivial paramagnet after a many-body basis rotation.
\begin{equation}
    U = \sum_{\{g_i\}} \left[ \prod_{\Delta^d} \nu_{d+1}(0,g_{i_1},\ldots g_{i_d})^{\epsilon(\Delta^d)} \right] \ket{\{g_i\}} \bra{\{g_i\}}\,.
\end{equation}
This trivial paramagnet can be realized in a local stoquastic Hamiltonian, which is the generalization of the $\mathbb{Z}_2$ paramagnet $-\sum_j \sigma^x_j$ to a generic group. Such a Hamiltonian is defined using group-valued qubits in Appendix~\ref{app:gauging}. Hence the conjugated Hamiltonian,  which has the SPT ground state, will also be SPF despite being non-stoquastic. 
Since $U$ assigns phase factors determined by evaluating local simplices, it can be expressed in terms of FDLU gates and the conjugated Hamiltonian remains local. Such Hamiltonians have been written down explicitly in~\cite{chen2013a}.

Our result is robust to perturbations away from this exactly-solvable limit: any stoquastic deformations of the stoquastic dual yields a deformed SPT Hamiltonian which remains SPF. For the Levin-Gu model, adding a nearest-neighbor Ising interaction $J \sum_{\langle i j \rangle} \sigma^z_i \sigma^z_j$ to the dual preserves stoquasticity. This perturbation is invariant under $U$, so these Ising interactions can also be added to the Levin-Gu Hamiltonian while preserving the SPF nature. Equivalently, perturbations which are diagonal in the computational basis cannot affect the sign structure of the SSE as any non-trivial sign can be absorbed by an overall constant shift in the Hamiltonian, so generic $\sigma^z$ perturbations can be freely added to the Levin-Gu Hamiltonian. The generalization to an arbitrary group is straightforward. These perturbations will generically drive the system into a symmetry-broken phase, with the critical theory described by a \textit{gapless} SPT state~\cite{scaffidi2017}. Alternatively, the $\mathbb{Z}_2$ symmetry can be broken by an external field $h \sum_j \sigma^z_j$. Varying the sign and spatial structure of these $\sigma^z$ terms produces a range of symmetry-broken phases distinguished by lattice space-group action. 

This model also admits SPF \textit{off-diagonal} perturbations; namely, one can tune between the Levin-Gu SPT and the trivial paramagnet, $H = (1-\alpha) H_{\text{Levin-Gu}} + \alpha H_{\text{trivial}}$. This is because $H$ is stoquastic for $\alpha < 0.5$, and the many-body basis rotation $U=(-1)^{N_{dw}}$ maps $\alpha$ to $1-\alpha$. Such a model was studied numerically for $\alpha < 0.5$ in~\cite{dupont2021}, with the $\alpha > 0.5$ behavior deduced via this duality transformation. Our claim is that no transformation is required - the $\alpha > 0.5$ region is SPF in its own right. Again, this construction is easily generalized to arbitrary groups.
\section{Gauging preserves sign structure}
We have demonstrated that every bosonic SPT in the group cohomology classification for unitary symmetries is SPF \textbf{in its local basis}. We now gauge these SPTs and obtain SPF Hamiltonians that are not locally connected to their stoquastic counterparts.
Gauging an SPT yields a twisted quantum double. For concreteness, we first gauge the Levin-Gu model and verify that SPF is preserved. Following~\cite{levin2012}, gauging the $\mathbb{Z}_2$ symmetry of both the Ising paramagnet and the Levin-Gu model produces the toric code and double semion model, respectively. We place qubits $\tau_{jk}$ on lattice edges and insert strings of $\tau^z$ operators connecting sites $j$ and $k$ in operators $\sigma^z_j \sigma^z_k$; for nearest-neighbor sites, this simply yields the operator $\sigma^z_j \tau^z_{jk} \sigma^z_k$.
We energetically penalize gauge flux with the term $\sum_{\triangle_{jkl}}\tau^z_{j k} \tau^z_{k l} \tau^z_{l j}$. The Gauss law $G_j \equiv \sigma^x_j \prod_{\langle j k \rangle} \tau^x_{j k} = 1$ is energetically imposed by a term $-\sum_j G_j$. 
\begin{equation}
\begin{aligned}
H_{\text{DS}} &= -\sum_j \tilde{A}_j + \sum_{\langle j k l \rangle} \tau^z_{jk} \tau^z_{kl} \tau^z_{lj} - \sum_J G_j \,,
\\
H_{\text{TC}} &= -\sum_j \sigma^x_j + \sum_{\langle j k l \rangle} \tau^z_{jk} \tau^z_{kl} \tau^z_{lj} -\sum_J G_j \,,
\\
\tilde{A_j} &\equiv \sigma^x_j \prod_{\triangle_{j k l}} i^{\frac{1}{2} (-1 - \sigma^z_k \tau^z_{kl}\sigma^z_l)} 
\label{eq:tqd}
\end{aligned}
\end{equation}The crux of our argument involves how the equality $H_{\text{Levin-Gu}} = U H_{\text{trivial}} U^\dagger$ behaves under gauging. Although $U$ cannot be implemented with symmetric FDLUs, the full unitary is symmetric and hence can be gauged, $U \rightarrow \tilde{U}$. An explicit construction of this gauging procedure is described in Appendix~\ref{app:gauging}, where we verify that $\tilde{U}$ remains a diagonal unitary transformation.
Now we must ask whether the equality $H_{\text{SPT}} = U H_{\text{trivial}} U^{\dagger}$ remains true for their gauged counterparts. This is only true if all strings of $\tau^z$ operators are assigned consistently, i.e. paths $\mathcal{P}_{jk}$ connecting sites $j$ and $k$ satisfy $\mathcal{P}_{jk} \mathcal{P}_{kl} = \mathcal{P}_{jl}$. The standard minimal coupling method of gauging, where nearest-neighbor sites are connected via their connecting edge, violates this rule. 

This argument shows that the toric code is related via many-body basis rotations to a non-local version of the double semion model, where $\sigma^z_j \sigma^z_k$ operators are attached to a generically non-local product of $\tau^z$ operators along a path $\mathcal{P}_{jk}$. We note that if we restrict ourselves to a subspace where gauge flux excitations are not present, then we can arbitrarily deform these paths and recover the conventional double semion model. Hence, the double semion model is SPF in this limit where we disallow non-trivial gauge flux. This is a physically-sensible limit that corresponds to an infinitely-large energy cost for gauge flux excitations, so we stress that this alone constitutes a counterexample to the claim of an intrinsic sign problem in the double semion model. 

Nevertheless, we show that even if we allow for gauge flux configurations, the double semion model remains SPF. We show in Appendix~\ref{app:cleaning} that the sign structure of the SSE is invariant under the choice of gauging path, thereby proving that any non-trivial sign will vanish in an SSE simulation of the double semion model. We note a technical point here, which is that in order for this additional unitary transformation to function, we must supplement the operators $A_j$ and $\sigma_j^x$ with projection operators $P_j \equiv \sum_{\triangle_{jkl}} \left( 1- \tau^z_{jk}\tau^z_{kl}\tau^z_{lj}\right)$, such that they only act when the flux on neighboring plaquettes is zero. This is a common convention used, for example, in~\cite{levin2012}, and is necessary to preserve Hermiticity of the Hamiltonian. This does not change the universal behavior of the system.



\begin{figure}[h]
  \centering
  \includegraphics[width=0.45\textwidth]{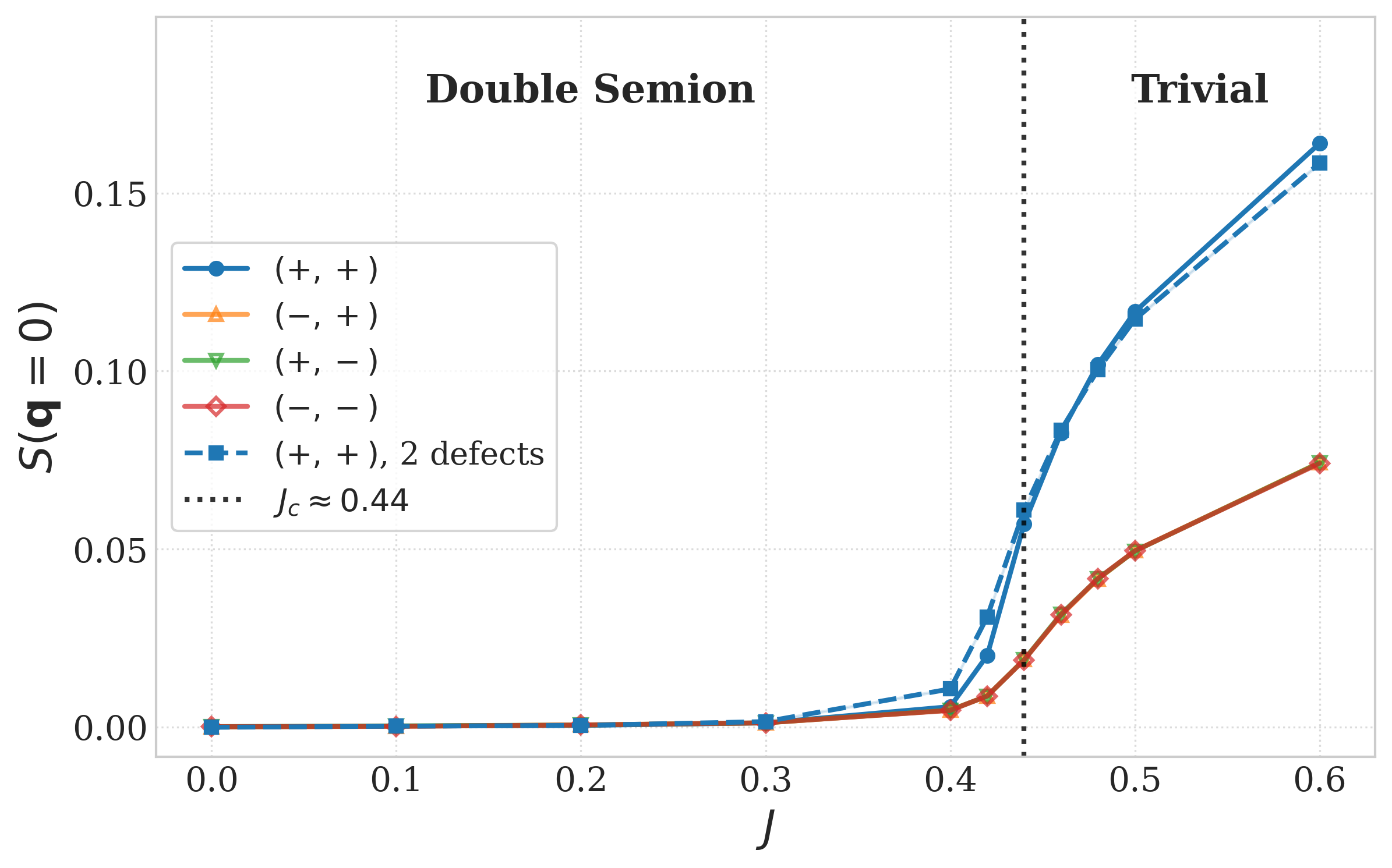}
  \caption{The condensation of semion/anti-semion pairs in the double semion model can be measured with a sign problem-free stochastic series expansion. Simulations were conducted on a $16 \times 16$ lattice at inverse temperature $\beta = 20$. Due to known ergodicity issues in simulations of gauge theories, we run separate QMC simulations in sectors with different magnetic fluxes, both global (specified by $(\pm, \pm)$ indicating absence or presence of flux in the $x$ and $y$ direction) as well as local magnetic flux excitations. Note that the three sectors $(+, -)$, $(-, +)$, and $(-, -)$ are related on the triangular lattice by a $C_3$ rotation, and the expectation values in these three sectors are nearly identical.}
  \label{fig:numerics}
  
\end{figure}

To demonstrate the validity of our claim, we conduct a QMC simulation of the double semion Hamiltonian in the presence of a ferromagnetic Ising interaction $-J \sum_{\langle j k \rangle} \sigma^z_j \tau^z_{jk} \sigma^z_k$. The Hamiltonian is no longer exactly solvable with this perturbation, but as we argued previously, diagonal perturbations do not affect the sign structure of the QMC. As shown in Fig.~\ref{fig:numerics}, we find a topological phase transition at $J_c \approx 0.4$, where semion/anti-semion pairs condense and lead to a trivial state. This condensation can be detected by long-range order in the gauge-invariant observable $S_{jk} \equiv \sigma^z_j \left[\prod_{\langle mn \rangle \in \mathcal{P}_{mn}} \tau^z_{kl} \right]\sigma^z_k$, or a peak in its Fourier transform $S(\vb{q}=0)$. Further details on the simulation and additional data are provided in Appendix~\ref{app:numerics}. One point to note is that models of this type are known to exhibit ergodicity issues in SSE simulations due to the difficulty in transitioning between sectors with different magnetic flux (through both contractible and non-contractible loops)~\cite{wu2023}. Such obstacles are likely circumventable with more sophisticated global updates, which are beyond the scope of this work. For the purpose of verifying the absence of a sign problem, we simply run independent SSE simulations within different magnetic flux sectors.

This procedure can be generalized to the gauging of any discrete SPT. However, properly defining the gauged unitary transformation for a generic group involves rather complex analysis of group cohomology cocycles. In the next section, we provide an alternative demonstration of the vanishing sign problem for the twisted quantum double of a generic group.
\section{Explicit sign problem-free models}
\label{sec:explicit}
We show that every twisted quantum double possesses an SPF model using the exactly-solvable models of twisted quantum doubles given in~\cite{hu2013a}. These models are equivalent to the ones studied in the previous sections by projecting out the matter fields. Because of the exact solvability, we can explicitly evaluate every term that appears in the SSE expansion in Eq.~\ref{eq:sse} and verify their positivity. The lack of a sign problem is not restricted to this exactly solvable limit, as any diagonal perturbations will preserve the sign structure.

We provide a self-contained definition in Appendix~\ref{app:tqd} and summarize here. The models live on an arbitrary triangulation of a two-dimensional surface, with group element-valued gauge fields $g_{jk}$ living on the bonds between sites $j$ and $k$. The sites are enumerated so that each bond has a definite orientation which points from site $k$ to $j$ if $j < k$ and the opposite direction otherwise. The Hamiltonian is specified by an inhomogeneous 3-cocycle $\omega$ and has two terms:
\begin{itemize}
    \item An operator $B_p$ on each plaquette that energetically imposes flatness of the gauge field.
    \item An operator $A_v^{g}$ for each vertex $v$ and group element $g$, which acts on all neighboring edges with $g$ and attaches a phase determined by the cocycle $\omega$. The precise phase factor is defined in Appendix~\ref{app:tqd}.
\end{itemize}
Analogous to~\cite{levin2012} and in the previous section, we dress $A_v^{g}$ with a projector which ensures all neighboring plaquettes are flux-free. 
Consider a generic term in the QMC expansion of Eq.~\ref{eq:sse}. Since the $B_p$ operators are diagonal in the computational basis, they do not modify the sign structure of the SSE, and it suffices to analyze strings of $A_v^g$. By construction, operators on different vertices commute and on the same vertex obey  $A^g_{v} A^h_v = A^{g h }_v$. Thus, any string reduces to one operator per site, $\bra{a} \prod_j A^{g_j}_j \ket{a}$. This contains the possibility of no operator acting on a site, which we represent by $g_j = 1$. 

Which choices of $\{g_j\}$ return $\ket{a}$ to itself? The crucial point underlying the SPF property is that \textit{the number of distinct combinations is extremely small}. Under $\prod_j A^{g_j}_j$, the gauge field $g_{jk}$ changes by $g \rightarrow g_j gg_k^{-1}$ (with directionality fixed by $j < k$). Requiring that this action on all bonds recovers the original state is a stringent condition on $\{g_j\}$. 

First, any non-trivial cycle must live in the flux-free subspace. If $\ket{a}$ has non-zero flux on a plaquette, then $A_v^g$ cannot act on the adjacent vertices. Gauge fields connected to those vertices now transform by a single group element, $g \rightarrow g_j g$, which forces $g_j = 1$. Applying this argument recursively implies $g_j = 1$ for \textit{all} $j$.

Now restrict to a locally flux-free subspace. Non-trivial cycles arise as follows. First, fix the operator on an arbitrary site $j$ to correspond to $g_j$. From this choice, all other operators are specified by the ``parallel transport'' $g_k = g_{kj}^{-1} g_j g_{kj}$. The flux-free condition ensures that this construction is path-independent, although the presence of non-contractible loops requires that $g_j$ belong to the centralizer subgroup of the two fluxes generated by computing $g_{jk}$ along the two non-contractible loops. These are the only non-trivial cycles in the QMC expansion; however, the action of this combined operator $\prod_j A_j^{g_j}$ on a closed manifold is the identity, as it acts as a global gauge transformation~\cite{hu2013a}. This relies on 3-cocycle identities and is shown in Appendix~\ref{app:tqd}. Therefore, the phase factor of these cycles is $1$ and the twisted quantum double is SPF.
\section{Conclusion}
We have shown that paradigmatic models of twisted quantum doubles, despite common belief, do not possess a sign problem when simulated with a conventional stochastic series expansion. This demonstrates that the ``sign problem'' the stochastic series expansion is sensitive to is fundamentally unrelated to wavefunction positivity. A recent work~\cite{seo2025a} argued - using stoquasticity as a proxy - that 398 of 405 bosonic topological orders in $d=2$ up to rank 12 have an intrinsic sign problem,. Within the space of low-rank topological orders, twisted quantum doubles occupy a small subset; below rank $12$, the only groups are $\mathbb{Z}_2$, $S_3$, and $\mathbb{Z}_3$. Their twisted variants yields $8$ counterexamples to this criteria. However, ~\cite{ringel2017} shows that every twisted quantum double of $\mathcal{G} = \mathbb{Z}_k$ possesses an obstruction to a stoquastic realization, so our construction provides a large class of counterexamples. The more general interplay between twisted quantum doubles and stoquasticity remains unexplored. We emphasize that our construction applies in dimensions other than $2$, with higher-dimensional quantum doubles supporting loop-like excitations~\cite{wang2014, wan2015}.

Design principles for non-stoquastic SPF Hamiltonians remain an open question. Although wavefunction non-positivity does not forbid SPF realizations, it does preclude stoquasticity. Thus in order to find these phases in, say, a more realistic spin model, such a Hamiltonian must be non-stoquastic and violate existing design principles. Our approach of starting with a stoquastic ``seed'' Hamiltonian and rotating to a non-stoquastic one may be generalizable to more conventional models, such as Heisenberg models. We also emphasize that diagonal perturbations leave the SSE sign structure unchanged while driving nontrivial physics away from exact solvability, providing another route to non-stoquastic SPF Hamiltonians.

Our results suggest a more intricate interplay between topological order and the sign problem. Although we have demonstrated the non-necessity of wavefunction positivity and our results are not fine-tuned to exactly solvable Hamiltonians, our results do rely heavily on the assumption that the gauge fluxes are frozen, i.e. no term in the Hamiltonian induces a transition to different flux sectors. It is unclear whether this constraint is fundamental; if so, it would imply that while there are still intrinsic obstacles to realizing twisted quantum doubles in QMC, these obstacles are more   

A promising direction to explore is the interplay between the sign problem and TQFT data in string-net models~\cite{levin2005}, which give a more general method for constructing TQFTs than gauging SPT states of discrete groups. The correspondence between the TQD models in Section~\ref{sec:explicit} and string-nets shown in~\cite{hu2013a} implies that string-net models of TQDs are SPF. However, string-net models give a platform to study the sign problem in TQFTs beyond group-like fusion categories.

Our results regarding the sign problem in SPT phases may potentially be generalizable beyond group cohomology constructions, in particular for continuous groups. The bosonic integer quantum Hall state~\cite{lu2012,senthil2013} is a prototypical example of an SPT in 2+1D protected by a $\UU(1)$ symmetry, for which no known SPF realization is known (we mention a related study of an $\UU(1) \times \UU(1)$ model which is SPF after a non-local duality transformation~\cite{geraedts2013}). The possibility of enriching the topological order with additional global symmetries~\cite{barkeshli2019} and searching for SPF models that can access phase transitions to symmetry-broken phases, is also a promising direction for future research.

There are intriguing open questions regarding the connection between properties of TQFTs and the sign problem, in particular regarding edge gappability. It was previously argued~\cite{seo2025a} that edge gappability is a necessary condition for the absence of an intrinsic sign problem. As we have stressed, this condition is not necessary for an SPF Hamiltonian, but rather a stoquastic one. However, because Abelian string-net models realize all non-chiral Abelian topological orders with a gappable edge~\cite{lin2014a}, our findings support the converse statement: for Abelian non-chiral topological orders, the presence of a gappable edge is a \textit{sufficient} condition for a SPF Hamiltonian. An open question is whether this condition is also necessary given our more robust definition of a sign problem. 
\section{Acknowledgements}
I thank Patrick Ledwith, Senthil Todadri, Sal Pace, Arkya Chatterjee, Carolyn Zhang, and Rahul Sahay for helpful discussions. This work was supported by the MIT Pappalardo Fellowship.
\appendix
\section{Gauging the trivializing unitary}
\label{app:gauging}
Here, we provide an explicit construction of the SPT entangler $U$ and its gauged version $\tilde{U}$. Since it is known that $U$ is a symmetric operator~\cite{chen2013a}, it must be possible to gauge it, although such a procedure is not obvious as $U$ is not locally symmetric. Or explicit gauging procedure here verifies that $U$ remains diagonal in the computational basis upon gauging, and demonstrates the manifestly non-local nature of the gauged operator $\tilde{U}$.

To facilitate this construction, we introduce the notion of of group-based qudits following~\cite{brell2015,pace2025}, which generalize qubits to an arbitrary finite group $\mathcal{G}$. On a lattice of arbitrary dimension with $N$ sites, each site $j$ carries a degree of freedom labeled by $g \in \mathcal{G}$. Basis states in the computational basis are $\ket{\{g_j\}} = \ket{g_1, g_2,\ldots, g_N}$. These transform in the regular representation of the group $\mathcal{G}$. The group-based $X$ operators are defined by
\begin{equation}
    \begin{aligned}
        \overrightarrow{X}_j^{(g)} &= \sum_{\{\bm{h}\} }\ket{g_j \bm{h}}\bra{\bm{h}}
      \\
     \overleftarrow{X}_j^{(g)} &= \sum_{\{\bm{h}\}} \ket{ \bm{h}\bar{g}_j}\bra{\bm{h}}
    \end{aligned}
\end{equation}and implement left/right multiplication with $\bar{g} \equiv g^{-1}$. The $Z$ operator is defined with an irrep of $\mathcal{G}$, $\Gamma$, and a choice of matrix element $(\alpha, \beta)$:
\begin{equation}
    \left[Z_j^{(\Gamma)}\right]_{\alpha \beta} = \sum_{\{\bm{h}\} }\left[\Gamma(h_j)_{\alpha \beta}\right] \ket{ \bm{h}}\bra{\bm{h}}
\end{equation}
In this notation, a trivial paramagnetic state $\ket{\psi} = \sum_{\{ g _i\}} \ket{\{g_i\}}$ is the ground state of the stoquastic Hamiltonian $H_{\text{para}} = -\sum_{j, h} \overrightarrow{X}^{(h)}_j$.

We will gauge these models by introducing group-valued degrees of freedom on the edges $\langle j k \rangle$. Each edge comes with an orientation, and we define $\overrightarrow{X}_{jk}^{(g)}$ as $\overrightarrow{X}$ if the edge $\langle j k\rangle$ points from $j$ to $k$, and $\overleftarrow{X}$ otherwise. Similarly, $Z_{jk}^{(\Gamma)}$ acts with the representation $\Gamma$ for the first case, and $\bar{\Gamma}(g) \equiv \Gamma(g^{-1})$ for the second.
We now construct the operator $U$ with these group-valued qubits. Recall that the operator $U$, when acting on a state $\ket{g_1,g_2,\ldots g_N}$, assigns a phase factor $\phi(g_1, g_2, \ldots g_N)$ determined by the SPT cocycles in Eq.~\ref{eq:cocycle}. The phase factor is invariant under a global symmetry action, so without loss of generality we consider the case where the group element on a site $j=1$ is equal to the identity. We now show that we can deduce the configuration on the rest of the system by measuring bond operators; by doing this, the desired phase factor can be in principle implemented. On these states, $\left[ Z_1^{\bar{\Gamma}} Z_k^\Gamma \right]_{\alpha\beta} = \Gamma(g_k)_{\alpha\beta}$. Using these matrix elements, we can construct the operator
\begin{equation}
  \frac{1}{\abs{\mathcal{G}}} \sum_{\Gamma} \Tr \left[\Gamma^\Gamma(h^{-1}) Z_1^{\bar{\Gamma}} Z_k^\Gamma   \right] = \delta_{g_k, h}\,,
\end{equation}
which follows from the fact that $\frac{1}{\abs{G}} \sum_{\Gamma} \Tr \Gamma(g) = \delta_{g, 1}$.
From this, we can write
\begin{equation}
    U = \sum_{h_2, h_3,\ldots , h_N} \left( \prod_{k \neq 1} \delta_{g_k, h_k} \right) e^{i \phi(1, h_2, \ldots , h_N)}\,.
\end{equation}
This yields the correct phase on all states $\ket{1, g_2, \ldots , g_N}$. Since $U$ is by construction symmetric, it will yield the same result on any state $\ket{g, g g_2, \ldots , g g_N}$.

We are now in a position to explicitly gauge the $U$. We do this by the minimal coupling
\begin{equation}
    \left[ Z_j^{\bar{\Gamma}} Z_k^\Gamma \right]_{\alpha\beta} \rightarrow \left[ Z_j^{\bar{\Gamma}} \left(\prod_{\langle m n \rangle \in \mathcal{P}_{jk}} Z_{mn}^\Gamma \right)Z_k^\Gamma \right]_{\alpha\beta}
\end{equation}
where $\mathcal{P}_{jk}$ is a path on the lattice that connects sites $j$ and $k$, and $\prod_{\langle m n \rangle \in \mathcal{P}_{jk}} Z_{mn}^\Gamma$ is a product of gauge field operators living on the links of the path connecting $j$ to $k$. This gauged operator is clearly still diagonal, as it is purely composed of $Z$ operators. We can also confirm that unitarity is conserved. To do this, consider the action of the gauged $U$ on a basis state. With these modifications, the operator ``$\delta_{g_k, h}$'' no longer evaluates to $1$ when $g_k = h$, but rather when $\left(\prod_{\langle m n \rangle \in \mathcal{P}_{1k}} g_{mn} \right) g_k = h$. Importantly, the gauged $U$ still assigns a well-defined phase to every basis state, and hence remains unitary. It is clear that this operator is highly non-local, as required by the obstruction to relating different topological orders by local unitaries. However, we stress that this is irrelevant to our arguments; one only needs the fact that the gauged $U$ is unitary and diagonal.

\section{Recovering the untwisted quantum double}
\label{app:cleaning}
As we have shown in the main text, the gauged unitary transformation maps the toric code to a non-local version of the double semion model $\tilde{H}_{\text{DS}}$, where rather than the minimal coupling prescription $\sigma^z_j \tau^z_{jk} \sigma^z_k$, we have a more generic path $\sigma^z_j \left[\prod_{\langle mn \rangle \in \mathcal{P}_{jk}} \tau^z_{mn} \right] \sigma^z_k \equiv \sigma_j^z \tau^z_{\mathcal{P}_{jk}} \sigma_k^z$ where the paths $\mathcal{P}_{jk}$ satisfy $\mathcal{P}_{jk} \mathcal{P}_{kl} = \mathcal{P}_{jl}$. We now show that this deformed double semion model has a sign structure identical to the ordinary one. Our approach in doing so is conceptually slightly different than the methodology in the main text. Here, we exploit the fact that the sign structure of the SSE is invariant under a relabeling of basis states; in other words, two Hamiltonians that map into each other under conjugation by a permutation matrix must clearly have the same SSE sign structure, as this only corresponds to a change in convention in how one labels the basis states. This relabeling maps the deformed double semion model to the standard one.

The deformed double semion model can be rewritten in a form which resembles the usual one, except with phase factors given by $i^{\frac{1}{2} (1 - \mathcal{W}_{kl}\sigma_k^z \tau^z_{kl} \sigma^z_l)}$, where $\mathcal{W}_{kl} \equiv \tau^z_{\mathcal{P}_{kl}} \tau^z_{kl}$ is the closed loop obtained by joining the path $\mathcal{P}_{kl}$ to the nearest-neighbor edge $\langle k l \rangle$. In a particular gauge sector, the effect of this term is to pick out a collection of ``defective'' bonds where $\mathcal{W}_{kl} = -1$ and the phase factor differs from the usual double semion model. Any configuration of these defects must obey the condition $\mathcal{W}_{jk}\mathcal{W}_{kl}\mathcal{W}_{lj} = 1$, i.e. the product around any site is equal to $1$. The fact that $ \mathcal{P}_{jk}\mathcal{P}_{kl}\mathcal{P}_{lj} =1$ follows from the definition of $\mathcal{P}$, and $\tau^z_{jk}\tau^z_{kl}\tau^z_{lj}=1$ due to the assumption that operators only act on locally flux-free subspaces. This condition means that these defective bonds can be identified with closed loops on the dual lattice, which in turn can be identified with gauge transformations. In other words, these defects would be cured if one was to perform the corresponding gauge transformation that acted $\textit{only}$ on the link degrees of freedom, $\tau^z_{jk} \rightarrow -\tau^z_{jk}$. We can think of this transformation as a relableling of basis states, $\ket{\alpha} \rightarrow \prod_{\mathcal{W}_{jk} = -1} \tau^x_{jk} \ket{\alpha} \equiv \ket{\tilde{\alpha}}$. This transformation commutes with both the Gauss law term and the magnetic flux term, so maps the deformed double semion model to the original. Hence, the two have the same sign structure.
\section{Exactly solvable models for twisted quantum doubles}
\label{app:tqd}
We provide a self-contained definition of the models defined in~\cite{hu2013a} for twisted quantum doubles. The models are defined on a two-dimensional graph consisting of only triangles, i.e. a triangulation of a two-dimensional surface. Group-valued degrees of freedom reside on the edges, with which we use the notation defined in Appendix~\ref{app:gauging}. The Hamiltonian is given by
\begin{equation}
    H = - \sum_v A_v - \sum_p B_p\,.
\end{equation}
The operator $B_p$ acts on each triangle and penalizes gauge flux excitations
\begin{equation}
    B_p = \frac{1}{\abs{\mathcal{G}}} \sum_\Gamma \Tr \left[ Z_{jk}^\Gamma Z_{kl}^\Gamma Z_{lj}^\Gamma \right]\,.
\end{equation}
Note that this term is diagonal in the computational basis and does not affect the sign structure of the QMC simulation.

\begin{figure}[ht]
    \centering
    \includegraphics[width=0.7\linewidth]{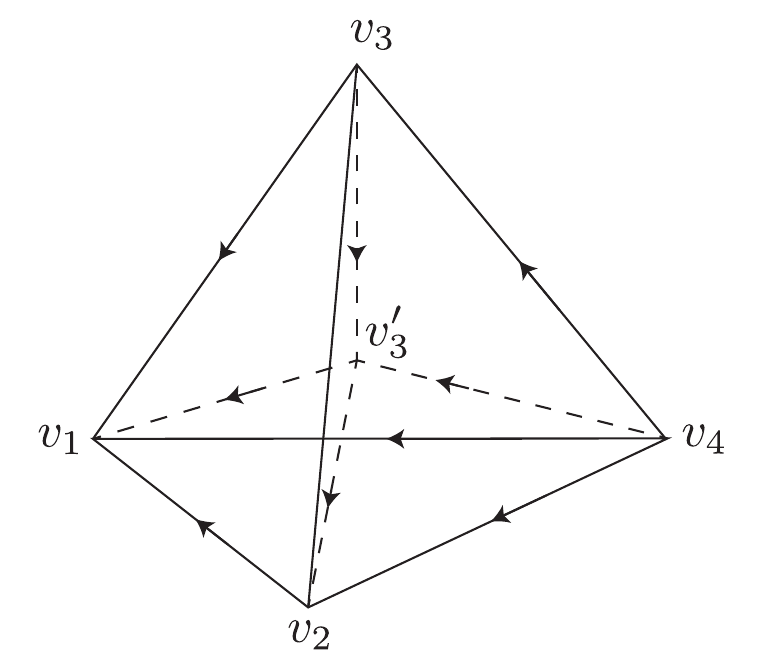}
    \caption{The action of $A_v^g$ on a vertex ($v_3$ in the figure) can be represented by inserting a new vertex $v_3'$ with $g_{v_3' v_3} = g$ and computing the phases associated to three of the plaquettes using the 3-cocycle $\omega$.}
    \label{fig:cocycles}
\end{figure}

The operator $A_v$ is given by a sum of terms for each group element
\begin{equation}
    A_v = \frac{1}{\abs{\mathcal{G}}} \sum_g A_v^g\,.
\end{equation}
$A_v^g$ acts on all edges connected to vertex $v$ with the group element $g$ (either left action by $g$ or right action by $\bar{g}$ depending on orientation) and attaches a phase determined by the 3-cocycle $\omega$. The phase is most easily described graphically through the intuition that $A_v^g$ evolves the graph in ``time'' by replacing the vertex $v$ by a new vertex $v'$, where the orientation of $v'$ is defined such that $v'$ is less than $v$ but greater than all other sites less than $v$. Using the vertex labeling in Fig.~\ref{fig:cocycles}, the action of $A^g_{v_3}$ comes with the phase
\begin{equation}
    \frac{\omega(g_{v_1 v_2}, g_{v_2 v_3'}, g_{v_3' v_3})\omega(g_{v_2 v_3'}, g_{v_3' v_3} ,g_{v_3 v_4})}{\omega(g_{v_1 v_3'}, g_{v_3' v_3}, g_{v_3 v_4})}
    \label{eq:cocycle}
\end{equation}
where
\begin{equation}
 \begin{aligned}
    g_{v_3' v_3} &= g\,,
    \\
     g_{v_1 v_3'} &= g_{v_1 v_3} \bar{g}\,,
     \\
    g_{v_2 v_3'} &= g_{v_2 v_3} \bar{g}\,,
    \\
    g_{v_3' v_4} &= g g_{v_3 v_4}\,.
    \label{eq:edgeRelabelings}
 \end{aligned}
\end{equation}
This phase can be obtained by combining the three phases associated with three tetrahedra that contain the edge $(v_3, v_3')$ via the 3-cocycle $\omega$.
The resulting state in the physical Hilbert space is obtained by setting $g_{v_1 v_3} = g_{v_1 v_3'}$ and likewise for $g_{v_2 v_3}$ and $g_{v_3 v_4}$.

\begin{figure*}[ht]
   \includegraphics[width=\textwidth]{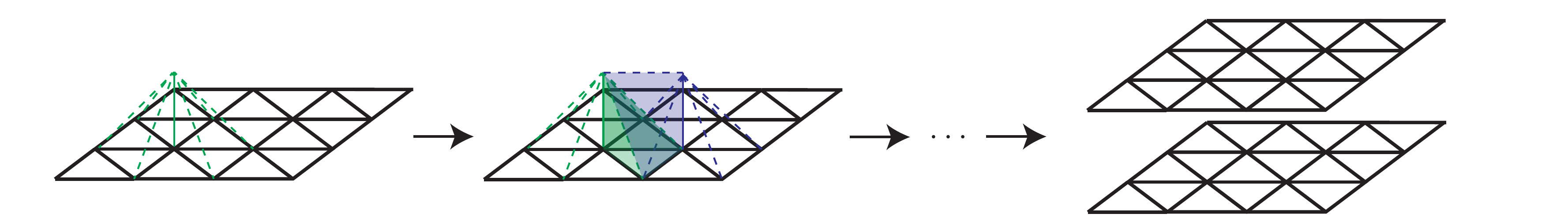}
   \label{fig:gaugeTransform}
   \caption{We illustrate the action of $A_v^g$ on neighboring sites. The action on neighboring vertices generates face-sharing tetrahedra whose contributions cancel out, leading to a trivial total action on a lattice without a boundary provided we choose our $A_v^g$'s such that the lattice is mapped back to itself under the transformation.}
\end{figure*}

With this interpretation, we can easily see how the global transformation in Section~\ref{sec:explicit}, where a vertex operator acts on each site with the group element determined by parallel transporting an initial element, is a trivial transformation on a closed manifold. The graphical depiction of this is shown in Fig.~\ref{fig:gaugeTransform}. 
Viewed in three-dimensional space, the action of $A_v^g$ introduces one tetrahedra for each edge connected to the vertex. The action of two of these transformations on neighboring vertices generates two tetrahedra that share a face. Acting on all vertices and assuming that our lattice does not have a boundary, we obtain a triangulation of a three-dimensional object whose inner and outer boundaries correspond to the original lattice and the transformed lattice, respectively. The factors from the internal plaquettes cancel out, and because the original and transformed lattice are the same by our choice of $A_v^g$, we find that the total phase is $1$.

\section{Numerical Simulations}
\label{app:numerics}

To numerically verify the lack of a sign problem in the double semion model, we conduct a SSE simulation of the double semion Hamiltonian in Eq.~\ref{eq:tqd} in the presence of a non-integrable Ising interaction. The full Hamiltonian is 

\begin{equation}
\begin{aligned}
  H_{\text{DS}} &= -\sum_j P_j \tilde{A}_j + \sum_{\langle j k l \rangle} \tau^z_{jk} \tau^z_{kl} \tau^z_{lj} 
  \\ & - \sum_J G_j  + J \sum_{\langle j k \rangle} \sigma^z_j \tau^z_{jk} \sigma^z_k\,,
  \\
  \tilde{A}_j &\equiv \sigma^x_j \prod_{\triangle_{j k l}} i^{\frac{1}{2} (-1 - \sigma^z_k \tau^z_{kl}\sigma^z_l)} \,,
  \\
  G_j &\equiv \sigma^x_j \prod_{\langle jk \rangle} \tau^x_{jk}\,,
  \\
  P_j &\equiv \sum_{\triangle_{jkl}} \left( 1- \tau^z_{jk}\tau^z_{kl}\tau^z_{lj}\right)\,.
  \label{eq:ds_app}
  \end{aligned}
\end{equation}

We first describe our procedure for running the simulation in the absence of a Gauss law term, and subsequrently explain the modifications caused by the Gauss law terms. Without any Gauss law terms, the gauge degrees of freedom are fully static. As such, the Hilbert space factorizes into sectors with definite gauge configurations. In each sector, we can replace $\tau^z_{jk}$ by its eigenvalues and the Hamiltonian $H_{\text{DS}}$ reduces down to a transverse-field Ising model with two key differences. The Ising interaction has sector-dependent couplings $\pm J$, and each transverse field term $\sigma^x_j$ has a phase factor $\prod_{\triangle_{jkl}} i^{\frac{1}{2} (-1 - \sigma^z_k \tau^z_{kl} \sigma^z_l)}$.

In a given sector, we can simulate this Hamiltonian using standard techniques for Ising-like models described in~\cite{sandvik2003}, including global loop updates. The double semion-like phase factor only causes a non-trivial modification to the SSE simulation \textit{if} it contributes a global phase. At each step in the SSE simulation, we evaluate this phase factor and confirm that it remains 1. 

We now turn on the Gauss law term. Since the Gauss law term commutes with every other term in the Hamiltonian, we can determine its effects analytically. Since the gauge fields are otherwise static, the combined action of all Gauss law terms in any matrix element must act trivially on the gauge fields in order for the matrix element $\bra{z} H_{b_1} H_{b_2} \ldots H_{b_n} \ket{z}$ to be non-vanishing. The only non-trivial combination of operators that accomplishes this action is $\prod_j G_j \equiv \prod_j \sigma^x_j$. Hence, the contributions to the SSE factorizes into two types of terms - ones where the collection of Gauss law terms act trivially, and ones where the Gauss law terms effectively impose ``twisted boundary conditions.'' The latter terms are only present in sectors with no local gauge flux, as the twisted boundary conditions demand that $\tilde{A}_j$ act on each site at least once, and local gauge flux prohibits this on neighboring sites. Within these sectors, we can proceed much like an ordinary SSE, but with an initial operator string given by $\prod_j \tilde{A}_j$ rather than the identity. 

\begin{figure}
    \centering
    \includegraphics[width=0.95\linewidth]{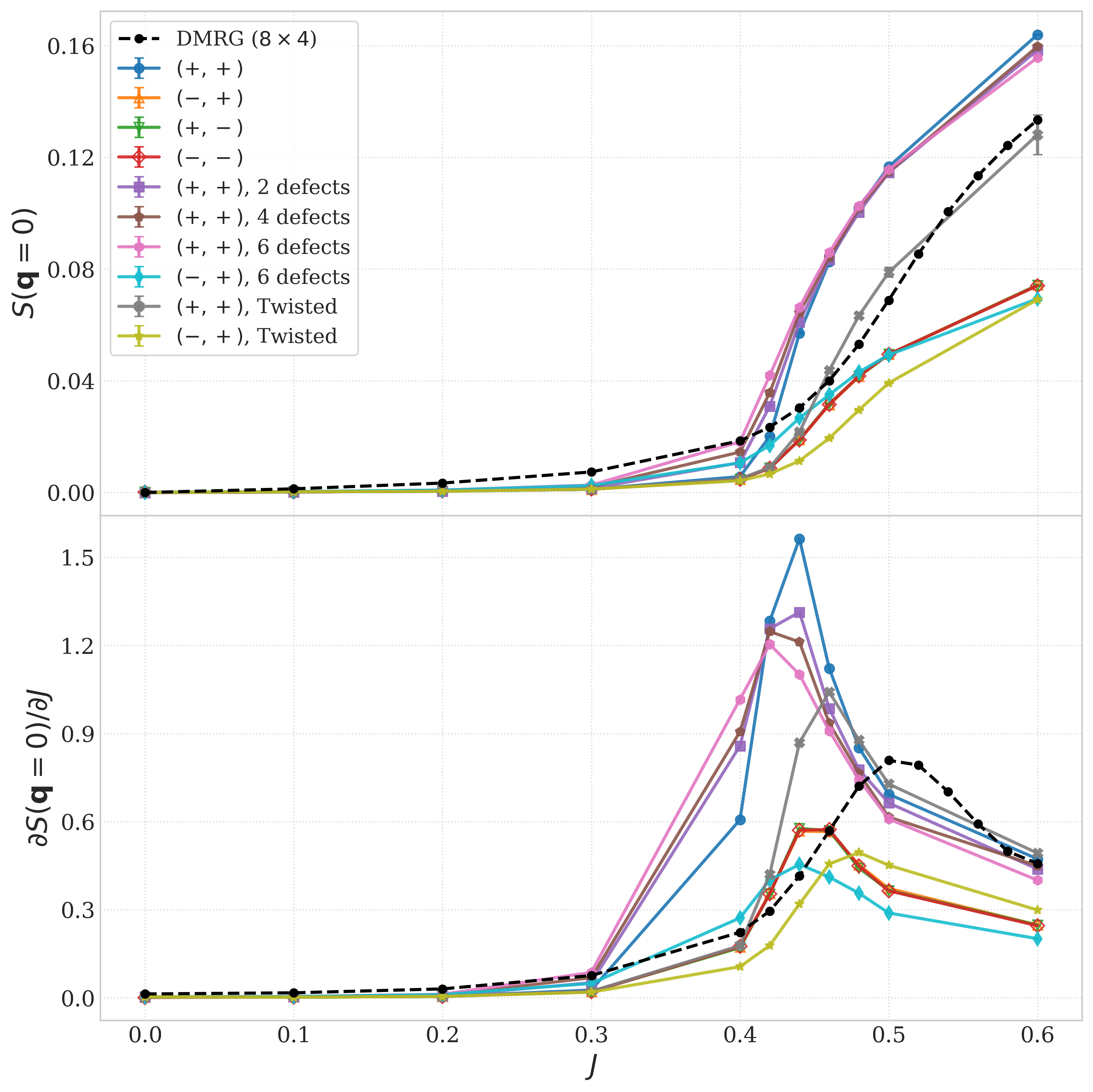}
    \caption{Across various magnetic flux sectors as well as with both twisted and untwisted boundary conditions, the double semion Hamiltonian in Eq.~\ref{eq:ds_app} remains sign problem-free and undergoes a phase transition into a trivial state at a critical value of Ising coupling $J_c \approx 0.44$. To confirm the simulation's validity, we also include DMRG simulations of the $(+, +)$ sector on an $8 \times 4$ open cylinder.}
    \label{fig:numerics_app}
\end{figure}

The simulations in this paper are all conducted over $10^5$ sweeps on a $16 \times 16$ lattice. We simulate the model at low temperatures $\beta = 16$ to detect the phase transition as plotted in Fig.~\ref{fig:numerics} and Fig.~\ref{fig:numerics_app} and have also run identical high-temperature simulations at $\beta=1$ to verify that the lack of a sign problem is robust across the entire Hilbert space.

\clearpage
\bibliography{refs.bib}
\end{document}